# Metal-coated nano-cylinder cavity for broadband non-classical light emission

I. S. Maksymov[1], M. Besbes[1], J.P. Hugonin[1], J. Yang[1], A. Beveratos[2], I. Sagnes[2], I. Robert-Philip[2], P. Lalanne*,[1]

[1] Laboratoire Charles Fabry de l'Institut d'Optique, CNRS, Univ Paris-Sud, Campus Polytechnique, RD 128, 91127 Palaiseau, France

[2] Laboratoire de Photonique et de Nanostructures, CNRS, Route de Nozay, 91640 Marcoussis Cedex, France

* Corresponding author: E-mail philippe.lalanne@institutoptique.fr

**Abstract:** A novel metal-coated nanocylinder-cavity architecture fully compatible with III-V GaInAs technology and benefiting from a broad spectral range enhancement of the local-density-of-states is proposed as an integrated source of non-classical light. Due to a judicious selection of the mode volume, the cavity combines good collection efficiency (≈45%), large Purcell factors (≈15) over a 80-nm spectral range, and a low sensitivity to inevitable spatial mismatches between the single emitter and the cavity mode. This represents a decisive step towards the implementation of reliable solid-state devices for the generation of entangled photon pairs at infrared wavelengths.

Solid-state cavity quantum electrodynamics offers a robust and scalable platform for quantum optics experiments and the development of quantum information processing devices. Solid-state sources of quantum states of light (single photons, indistinguishable or entangled photons) based on semiconductor quantum dots (QDs) embedded in microcavities have seen rapid progress during the past decade [1-3]. The microcavity (e.g. a photonic-crystal (PhC) cavity or a micropillar) induces an increase of the spontaneous emission (SE) rate and a preferential funneling of the emitted photons into the cavity mode opening the possibility of a high collection efficiency of the emitted photons. It additionally permits to circumvent some of the effects of dephasing and partially restores the coherence between the emitted photons [2,4]. However, in practice, microcavities impose a spectral and spatial matching of the emitting dipole with the cavity mode, which still limits the performance of such sources. Techniques are presently being developed to guaranty the matching, by deterministically localizing the cavity around the randomly-placed emitting dipole [5,6] or by deterministically localizing the emitting dipole [7]. Another approach consists in designing novel architectures that are naturally less sensitive to any spectral or spatial matching. Broadband and efficient single-mode-photon sources have been recently demonstrated without any cavity effect, by use of PhC waveguides [8-10] or nanowire antennas [11, 12]. While these single-quantum-emitter architectures may offer high multi-photon probability suppression [10-12] and record collection efficiency through a microscope objective [11], the coherence of the emitted light is weak, and because the phase memory decay of most solid-state emitters is much faster than their radiative recombination lifetimes [2] quantum properties such as indistinguishability or entanglement are not achievable. In this letter, we propose a new architecture, ensuring both a high collection efficiency and a large SE-rate enhancement (thus restoring coherence), while relaxing constraints on spectral and spatial matching. This is achieved with a nanocylinder antenna, which operates in an intermediate confinement regime with a mode volume $V \approx 0.002\lambda^3$. The mode volume is approximately 10-times larger than those achieved with ultrasmall plasmonic nanoantennas, such as nanogap in metallic dimers [13], relaxing the constraint on the emitting dipole position, but much smaller than those of PhC cavities. Moreover, the high Purcell factor is achieved over a broad 80-nm spectral range, relaxing the constraint on the spectral matching. Such geometry is consequently a promising candidate for enhancing the capability of solid-state sources of quantum states of light, such as indistinguishable or entangled photons.

Figure 1(a) sketches the geometry considered in this letter, a textbook case consisting in a semiconductor GaAs cylinder of radius *a* and height *H* coated by a thin dielectric $Si_3N_4$ adlayer and by a thick silver layer. The 5-nm dielectric layer is introduced for preventing non-radiative recombination at the interfaces and for increasing the bottom metallic mirror reflectance. In comparison with PhC cavities or microposts, the metallic coating offers smaller mode volumes and a suppression of off-resonant leaky modes [14], which boosts the emitting dipole-coupling with the cavity mode. The main drawback is the inevitable non-radiative decay rate induced by metallic losses. Except when otherwise specified, the emitting dipole is modeled as an in-plane on-axis unpolarized dipole **J** located at a distance $h_2$ from the top air-interface.

The design and interpretation are based on an approximate Fabry-Perot (FP) model. We start by considering the fundamental nanocylinder mode, denoted $TE_C$ hereafter and assumed to be formed in the FP picture by the bouncing back and forth of the guided $TE_{11}$ cylinder modes [15] between the top and bottom interfaces. The amplitudes $A^+$ and $A^-$ in Fig. 1(a) respectively represent the excitation coefficients of the upward and downward $TE_{11}$ modes [Fig. 1(b)], $r_t$ and $r_b$ are the modal reflection coefficients of the top and bottom interfaces, and $T(\theta)$ denotes the fraction of the energy carried by the $TE_{11}$ mode, which is scattered by the top interface into a cone defined by the opening angle $\theta$. Figure 1(c) shows the effective index $n_{eff}$ dependence of the TE$_{11}$ mode with the cylinder radius $a$ at $\lambda_0$=950 nm. We will operate above the TE11 mode cut-off radius of 50 nm. The second TM$_{01}$ mode possesses an antinode of the electric field on axis; it will be considered for analyzing the impact of off-axis emitting dipoles on the device performance. The Purcell factor $F_P$, defined as the normalized SE rate into the fundamental nanocavity mode TE$_C$, is given by [16]

$$F_P = (1-|u_t|^4 R_t)|A^+|^2 + (1-|u_b|^4 R_b)|A^-|^2, \quad (1)$$

where $R_b=|r_b|^2$, $R_t=|r_t|^2$, $u_b=\exp(ik_0 n_{eff} h_1)$ and $u_t=\exp(ik_0 n_{eff} h_2)$ see Fig. 1(a), with $k_0=2\pi/\lambda$, $A^+=A_s[1+r_b u_b^2]/[1-r_t r_b(u_b u_t)^2]$ and $A^-=A_s[1+r_t u_t^2]/[1-r_t r_b(u_b u_t)^2]$, $A_s$ being a dipole-field coupling coefficient proportional to the oscillator strength. The extraction efficiency $\eta(\theta)$ is defined as the fraction of emitted photons that are collected above the top interface for the opening angle $\theta$

$$\eta(\theta) = T(\theta)|u_t|^2|A^+|^2/P_T, \quad (2)$$

where $P_T=F_P + \gamma$ is the total SE rate and $\gamma$ is the decay rate into other cavity modes including the decay rate into the continuum of radiation modes of the cylinder. As will be shown later by comparing the FP predictions with fully vectorial finite-element-method (3D FEM) data, $\gamma$ is negligible for on-axis dipoles. This drastically simplifies the design and can be attributed to the beneficial contribution of the metal-coating that inhibit the dipole-field coupling into the continuum of off-resonant radiation modes, a remarkable phenomenon already observed in micropillars [14] and nanowire lasers [17].

The coefficients, $r_t$, $r_b$ and $T(\theta)$, of the TE$_{11}$ mode [Figs. 2(a) and 2(b)] are calculated with a 3D frequency fully-vectorial aperiodic-Fourier-Modal method (a-FMM) [18] for $\lambda_0$=950 nm. The frequency-dependent refractive index of silver is taken from [19], $n_{SiN}$=1.95 and $n_{GaAs}$=3.45. Above cutoff, the reflectance of the bottom interface is $R_b \approx 0.95$, independently of the cylinder radius. That of the top interface, $R_t$, rapidly varies at cutoff before slowly decreasing to reach the asymptotic value of $R_\infty=(n_{GaAs}-1)^2/(n_{GaAs}+1)^2=0.3$ for large radius ($a \to \infty$). It is considerably smaller than the mirror reflectance usually implemented in micropillars or PhC cavities. Nevertheless, the dipole-field coupling in the nanocavity is strengthened, due to the small modal volume, and high Purcell effects are achieved over a broad spectral range. The outcoupling efficiency $T(\theta)$ is approximately equal to $1-R_t$ for $\theta=\pi/2$. The energy loss, $1-R_t-T(\pi/2)$ is due to the launching and absorption of plasmons on the top metallic interface and reaches a maximum value of 0.1 for $a=100$ nm ($a \approx \lambda_0/10$ is the relevant value of the radius for the antenna) and further decreases with $a$; it is only 0.05 for $a=200$ nm.

In Fig. 2(c) and 2(d), we show the dependence of $F_P$ and $\eta$ with the cylinder radius $a$ for $\lambda_0=950$ nm. The results are obtained at resonance for a dipole source placed on the antinode of the cavity mode, implying that, as $a$ varies, the cylinder height $H$ and the dipole location $h_2$ also vary to maintain the FP resonance condition for the $TE_{11}$ mode (see the inset). The quantitative agreement between the FP model predictions (curves) and fully-vectorial data obtained with a 3D FEM software (circles) shows that the model accurately captures the physics governing the dipole-field coupling of the nanocavity and convincingly supports our assumption that $\gamma$ is negligible for on-axis dipoles. The general trends in Figs. 2(c) and 2(d) are understood by considering the different loss mechanisms, including absorption at the bottom and top interfaces, and absorption in the silver coating when the $TE_{11}$ mode bounces back and forth in the cylinder. With the FP model, we have analyzed their respective impacts, by artificially setting either $|R_b|=1$, $|R_t|=1-T(\pi/2)$ or $Im(n_{eff})=0$. We find that absorption in the silver cladding is dominant. As $a$ increases, this absorption decreases and this explains well why the efficiency increases with $a$ to reach a maximum value of $1-R_b$ for large $a$'s. The general trend for $F_p$ is primarily explained by mode-volume considerations. However, near cutoff, the beneficial effect of the volume reduction is balanced by the sudden increase of the nanocylinder height $H$ (due to a decrease of $n_{eff}$, see Fig. 1(c)) and of the concomitant loss in the silver cladding. $F_P$ reaches its maximum value for $a≈58$ nm and for a height of $H=152$ nm. For this combination of geometric parameters, the losses are still compensated by a strong dipole-field coupling due to a high reflectance of the top interface and to the small volume.

Figure 3 shows the spectral characteristics of the antenna for an intermediate cylinder radius, $a=100$ nm, which provides a good trade-off between extraction efficiencies ($\eta=44\%$ for $\theta=45°$) and Purcell factors ($F_P=15$). The striking feature is the broadband operation: the full-width at half-maximum of the $F_P$ spectrum in Fig. 3(a) is as large as the inhomogeneous broadening of self-assembled quantum-dot emission lines for instance and the extraction efficiency in Fig. 3(b) is virtually independent of the wavelength from 800 to 1100 nm. This spectral range is comparable to that recently reported for semiconductor nanowire antennas [11-12], but additionally, the present architecture supports a strong enhancement of the local density of states. The latter is essential for the generation of indistinguishable single photons and polarization entangled photons arising from the bi-exciton cascade of a single QD, so that photons emission occurs before any dephasing mechanisms take place [20,2,4]. In order to obtain a maximally entangled state, $|\Phi^+\rangle=(|HH\rangle+|VV\rangle)/\sqrt{2}$, it is additionally necessary that both horizontal (H) and vertical (V) polarizations undergo the same Purcell enhancement and that both exciton and bi-exciton emissions are collected identically [21].

These requirements are automatically satisfied for on-axis dipoles due to the invariance by rotational symmetry. In practice, however, the QD is not exactly on-axis, the degeneracy is lifted, and the degree of entanglement is inevitably lowered. As will be shown later, the fidelity of entangled photons pairs, i.e. the similarity of the emitted state to the maximally entangled Bell state $|\Phi^+\rangle=1/\sqrt{2}\,(|HH\rangle+|VV\rangle)$, can be accurately predicted for small misalignments ($a_0<50$ nm) with an extended FP model, by considering only two cavity modes, the $TE_C$ fundamental mode and an additional $TM_C$ mode formed by the bouncing of the $TM_{01}$ cylinder modes in the nanocylinder. $TM_{01}$ is responsible for the degeneracy lift at small misalignments; it has no azimuthal electric-field component and its radial component possesses an on-axis node [15], excluding its coupling with on-axis QDs. With the a-FMM, we have calculated the modal reflectance of the $TM_{01}$ mode at the top and bottom interfaces.

Figure 4a shows the *normalized* SE rate of radial and orthoradial dipoles as a function of the radial misalignment $a_0$. The results are obtained with the extended FP model for the selected nanohole geometry ($a$=100 nm, $H$=65 nm). We notice that, due to the coupling to the $TM_C$ mode, the recombination path of H-polarized photons (red-dotted line) possesses a Purcell factor higher than that of V-polarized photons (blue-dashed line); the H-polarized photons are facilitated with regard to the V-polarized ones, and this may lead to a non-maximally entangled state or even to a loss of entanglement. To quantify this eventuality, we calculate the fidelity as $Tr(\rho\Phi^+)$, where $Tr(.)$ is the trace and $\rho$ is the density matrix of the entangled photons emitted by an off-axis QD. The density matrix is estimated using the formalism in [21], under the assumption that the anisotropic spin exchange energy of the QD is negligible. The fidelity [Fig. 4(b)] is significantly higher than the threshold value of 0.85 required to violate the Bell inequalities, for off-axis displacements up to 60 nm. Furthermore, it remains larger than 90% over a remarkable range of off-axis locations, $0<a_0<45$ nm, an alignment precision largely within the capabilities of current fabrication technologies [6]. Note that the 3D FEM calculations of the total SE rate of the radial and orthoradial dipoles validate the bi-mode assumption of the FP model.

SE-rate asymmetry is not the only cause of disentanglement; path distinguishability may also arise from emission into different cavity modes with different radiation diagrams [21]. Because H-polarized photons couple to both $TE_C$ and $TM_C$ modes, whereas V-polarized photons only couple to the $TE_C$ mode, it is important to examine the asymmetry induced by the $TM_C$ mode. An important prediction of the extended FP model is that the $TM_{01}$ modal reflectance at the top interface is very large ($\approx$0.99 at $\lambda_0$=950 nm), even larger than the metallic reflectance at the bottom interface. The lifetime of the $TM_C$ mode is thus limited by Ohmic losses rather than by radiation into the far-field. Therefore the dipole-field coupling to the $TM_C$ mode is essentially non-radiative and the total ($\theta = 90°$) extraction efficiency $\eta_{TM}$ associated to the $TM_C$ mode is much smaller than that associated to the $TE_C$ mode. This is illustrated in Fig. 4c, which compares the probabilities, $P_{TE}$ and $P_{TM}$, of collecting into the first lens a photon originating either from the $TE_C$ (red-dotted curve) or from the $TM_C$ (blue-dashed curve) mode. $P_{TE}$ is equal to $\eta_{TE}\beta_{TE}$, where $\beta_{TE} = F_{P,TE}/(F_{P,TE}+F_{P,TM})$ is the β-factor of the $TE_C$ mode and $F_{P,TE}$ and $F_{P,TM}$ represent the Purcell factors of the $TE_C$ and $TM_C$ modes, respectively. With similar notations, we have $P_{TM} = \eta_{TM}\beta_{TM}$ and $\beta_{TM} = F_{P,TM}/(F_{P,TE}+F_{P,TM})$. Although the $TM_C$ mode gives rise to a strong Purcell effect (Fig. 4a), it is weakly affecting the far field collection, $P_{TE} \sim 10^3\ P_{TM}$ even for large misalignments. Since in addition the radiation diagrams of the $TM_C$ and $TE_C$ modes weakly overlap (Fig. 4d), the $TM_C$ contribution can be spatially filtered without diminishing the source's brightness, and we are confident that the fidelity is only limited by the path asymmetry of the Purcell effect (Fig. 4c).

In summary, we have investigated a new architecture for implementing III-V GaInAs solid-state sources of quantum states of light with large Purcell factors and with a good extraction efficiency over a remarkably large 80 nm spectral range. The proposed nanocavity is predicted to be capable of generating polarization entangled photon pairs with a high fidelity, even for QD misalignments up to 50 nm. The intermediate confinement regime ($V \approx 0.002\lambda^3$) of the investigated nanocavity completely relaxes the stringent spectral matching condition inherent to conventional PhC cavities, while keeping spatial matching at a reasonable level compatible with state-of-the-art nanofabrication facilities. This is not achievable with ultrasmall plasmonic antennas [13] and we believe that the present approach is a good compromise (see EPAPS) towards the integration of highly efficient solid-state sources of non-classical light.

**Acknowledgements** The work has been performed under the NanoEPR project of the 2006 NanoSci-ERA European program.

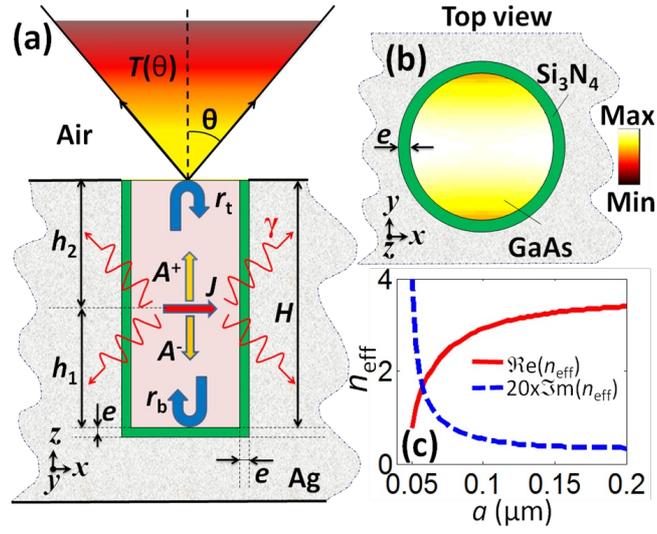

**Fig. 1** (Color online). (**a**) Cross-section of the metal-coated nanocylinder showing the main physical quantities involved in the FP model. (**b**) Top-view including the $|E_x|^2$ field profile of the fundamental $TE_{11}$ mode, calculated for $a$=100 nm, $e$=5 nm and $\lambda$=950 nm. (**c**) Effective index $n_{\it eff}$ of the $TE_{11}$ mode for $\lambda$=950 nm.

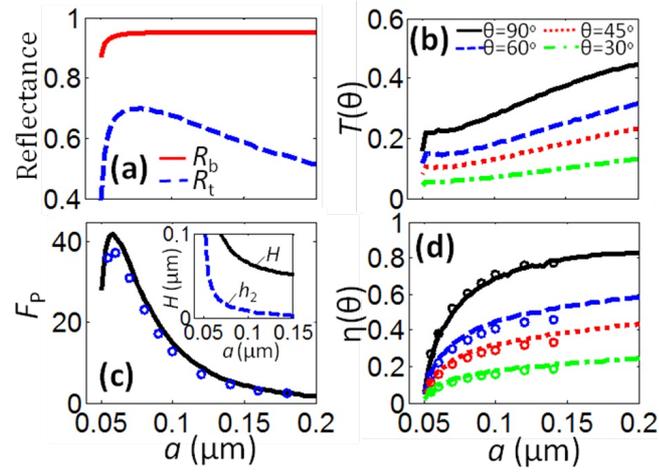

**Fig. 2** (Color online) Main physical quantities of the FP model as a function of the cylinder radius $a$ for $\lambda$=950 nm and $e$=5 nm. **(a)** Reflectance of the top and bottom interfaces. **(b)** Out-coupling efficiency of the fundamental mode. **(c)** Purcell factor. The inset shows the $H$ and $h_2$ that maintain the FP phase-matching resonance condition as $a$ varies. **(d)** Extraction efficiency for several θIn **(c)** and **(d)**, circles are obtained with the 3D FEM.

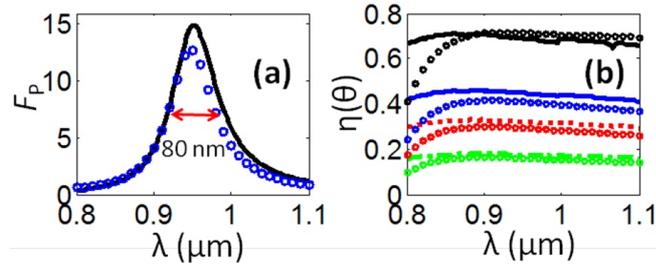

**Fig. 3** (Color online) Spectral performance of the nanocavity for $a$=100 nm, $e$=5 nm, $H$=65 nm and $[h_1,h_2]$=[57,8] nm. (**a**) Purcell factor. (**b**) Extraction efficiency for several $\theta$ (same markers and colors as in Figs. 2c and 2d). The efficiency drop predicted with the FEM at $\lambda$<850 nm is due to a resoannce of the $TE_{21}$ mode, near its cut-off at l=750 nm.

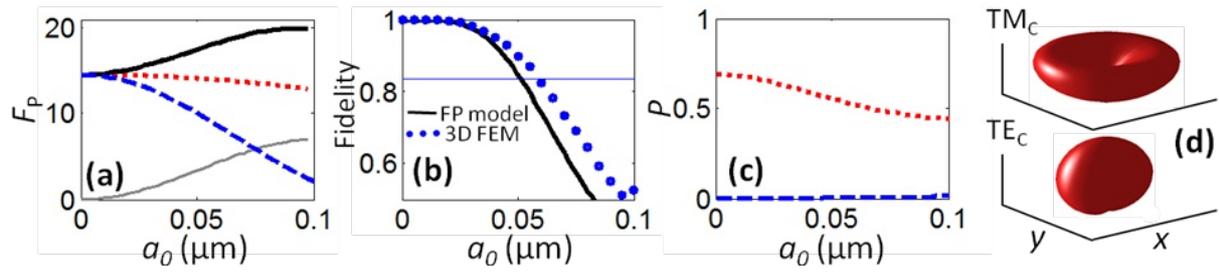

**Fig. 4** (Color online) Influence of the off-axis dipole location $a_0$ at $\lambda$=950 nm. **(a)** Red-dotted and blue-dashed curves: SE rate of the radial and orthoradial dipoles into the $TE_C$ mode. Thin gray-solid curve: SE rate of the radial dipole into the $TM_C$ mode. Black-solid curve: *total* SE rate of the radial dipole due to radiation into both the $TE_C$ and $TM_C$ modes. **(b)** Fidelity of entanglement (0.85 threshold is indicated). **(c)** Probability to collect a photon from the $TE_C$ (red-dotted) and $TM_C$ (blue-dashed) modes for $\theta$=90°. **(d)** Far-field radiation patterns of the $TE_C$ and $TM_C$ modes.